# Comment on "First principles calculation of lattice thermal conductivity in mono- and bi-layer graphene" by B. D. Kong, S. Paul, M. Buongiorno Nardelli and K. W. Kim

## A.A. Balandin, D.L. Nika and E.P. Pokatilov

*Nano-Device Laboratory[+], Department of Electrical Engineering, University of California – Riverside, Riverside, California 92521 USA*

## A.S. Askerov

*Department of Theoretical Physics, Moldova State University, Chisinau, MD-2009, Republic of Moldova*

*Abstract*

In a recent preprint Kong *et al*, arXiv: 0902.0642v1 (2009) claimed to calculate the lattice thermal conductivity of single and bi-layer graphene 'from first principles". The main findings were that the Umklapp-limited thermal conductivity is only slightly higher than that of high-quality bulk graphite along the basal plane, and that it does not strongly depend on the number of atomic layers. Here we explain that the calculation of Kong *et al* used a truncation procedure with a "hidden" parameter, a cut-off frequency for the long-wavelength acoustic phonons, which essentially determined the final result. Unlike in bulk graphite, there is no physical justification for introducing the cut-off frequency for the long wavelength phonons in graphene. It leads to substantial underestimation of graphene's lattice thermal conductivity and a wrong conclusion about the dependence on the number of atomic layers. We outline the proper way for calculating the lattice thermal conductivity of graphene, which requires an introduction of other scattering mechanisms to avoid a logarithmic divergence of the thermal conductivity integral.

---

[+] Web-site: http://ndl.ee.ucr.edu/

In a recent preprint Kong *et al* [1] performed the "first principle" calculation of the thermal conductivity of mono- and bi-layer graphene and found that it is around 2200 W/mK at 300 K. The authors also concluded that it does not depend significantly on the number of layers. The calculation used the phonon dispersion and the phonon-mode dependent Gruneisen parameters $\gamma(q)$ obtained with the help of QUANTUM-ESPRESSO package (where $q$ is the phonon wave vector). Here we explain that in order to obtain a finite value for the thermal conductivity of graphene, which is limited only by the phonon Umklapp scattering, Kong *et al* [1] had to use a truncation procedure with a "hidden" parameter – the cut-off frequency for the long-wavelength acoustic phonons. The selection of this cut-off frequency, which was not described in the text, effectively determined the final result. The physical reasoning previously used for introducing such a cut-off frequency in bulk graphite is not valid for the single layer graphene. For these reasons, both conclusions of Kong *et al* [1] do not have any merit.

Kong *et al* [1] limited their consideration to the three-phonon Umklapp scattering. For the scattering rate, $1/\tau_U$, they used a simplified expression given by Klemens for the basal plane of bulk graphite [2]:

$$\frac{1}{\tau_U} = 2\gamma^2 \frac{kT}{M\upsilon^2} \frac{\omega^2}{\omega_m}, \tag{1}$$

where $k$ is the Boltzmann constant, $T$ is the absolute temperature, $M$ is the mass of atoms, $\upsilon$ is the sound velocity, $\omega$ is the phonon frequency and $\omega_m$ is upper-bound cut-off frequency. Note that Kong *et al* [1] refer to $\omega_m$ as "Debye" frequency although it is defined differently from the actual Debye frequency. There is some ambiguity with $\omega_m$ but it is not of principle matter, and this parameter can be deduced from the known phonon dispersion of the material.

The problems start when Eq. (1) is substituted in the general expression for the thermal conductivity as the only scattering mechanism for phonons. It is known that in this case the integral for the thermal conductivity $K$ has a logarithmic divergence, which requires an introduction of the long-wavelength (low-frequency) cut-off frequency $\omega_C$ for acoustic phonons. In his seminal work for bulk graphite [2], Klemens derived an analytical expression for the Umklapp-limited thermal conductivity, which clearly shows that

$$K \propto \frac{1}{\omega_m} \int_{\omega_C}^{\omega_m} \frac{d\omega}{\omega} \propto \frac{1}{\omega_m} \ln\left(\frac{\omega_m}{\omega_C}\right). \quad (2)$$

As one can see from Eq. (2), an attempt to integrate over the whole Brillouin zone (BZ), starting from the zone center ($\omega(q=0)=0$ for acoustic phonons), requires setting $\omega_C=0$ and leads to infinite thermal conductivity. Klemens provided the physical reasoning for selection of $\omega_C$ for bulk graphite. He argued that at very low frequencies there appears coupling with the cross-plane phonon modes and heat starts to propagate in all directions, which reduces the contributions of these low-energy modes to heat transport along basal planes to negligible. The presence of the ZO' phonon branch near ~4 THz in the spectrum of bulk graphite has been used to support this argument and to introduce a corresponding cut-off frequency. The selection of $\omega_m/\omega_C$, an average Gruneisen parameter $\gamma=2$ and sound velocities, led to the calculated thermal conductivity of basal planes in bulk graphite of ~1900 W/mK, closely fitting the available experimental data [2].

In his preprint, Kong *et al* [1] did not specify the value of $\omega_C$ they used to obtain the thermal conductivity of the single and bi-layer graphene and did not discuss the truncation procedure for the integral at low frequency. They mentioned that *ZO'* branch appears in the bi-layer graphene. Here we would like to stress that the physical reasoning used for bulk graphite is completely without the merit in the case of graphene. The mode coupling at long wavelength does not lead to any heat transport in cross-plane direction in the system of one or two atomic planes. Moreover, *ZO'* mode is absent in the single layer graphene [4]. If Kong *et al* [1] followed the procedure applied to graphite consistently, they would have obtained infinite thermal conductivity for the single layer graphene (no *ZO'* mode; $\omega_C \to 0$). Indeed, one can get any value of the thermal conductivity by tacitly setting $\omega_C$. The conclusion about the dependence of the thermal conductivity on the number of atomic layers would also depend on the $\omega_C$ choice (rather than on the Gruneisen parameter).

Physically, the ambiguity with the approach limited to the three-phonon Umklapp scattering only is explained by the fact that this scattering mechanism becomes inefficient for small phonon wave vectors $q$. The interaction of two phonons with small $q$ does not result in another phonon with the wave vector outside of the 1$^{st}$ BZ. In real graphene flakes the long-wavelength phonon transport will be limited by other scattering mechanisms such as graphene edge scattering, point defect scattering, micro-crystalline size, etc. The inclusion of other relevant scattering mechanisms together with the Umklapp scattering in the thermal conductivity integral prevents divergence at low phonon frequencies $\omega$. The method, which includes other relevant scattering process, does not rely on any *ad hoc* truncation procedure and only requires some realistic parameters for the graphene flakes (e.g., characteristic size, defect densities, etc.). We have followed this method to calculate the thermal conductivity of a single layer graphene [3]. In our calculation we did not make use of the approximate Eq. (1) but rather calculated the three-phonon Umklapp scattering rates for the actual BZ of graphene directly accounting for all phonon transitions allowed by the momentum and energy conservation. It was found that the RT thermal conductivity of a single layer graphene is in the range 2000 – 5000 W/mK depending on the flake size, edge roughness and defect density [3]. This is substantially higher than the thermal conductivity of bulk graphite. The thermal conductivity calculated by us is in excellent agreement with the experimental works, which reported the near RT thermal conductivity in the range from ~3000 – 5000 W/mK for a set of single layer graphene flakes of different sizes [4-5].

It is interesting to note that in his later work (year 2000) on the subject [6], Klemens examined specifically the case of graphene (he actually used the term "graphene"). He stated directly that unlike in bulk graphite, in graphene "*the phonon gas is two-dimensional down to zero frequency*", and noted that the logarithmic divergence is prevented through limitations to the phonon mean free path from the linear dimensions of graphene sheets. This means that the introduction of any sort of cut-off frequency for graphene in the Umklapp scattering term, as apparently done by Kong *et al* [1] in their calculation, is arbitrary and unphysical. The thermal conductivity of single layer graphene calculated with the Umklapp scattering only goes to infinity due to the logarithmic divergence. Klemens' assertion that other scattering mechanisms are required in order to obtain a meaningful value for the thermal conductivity of graphene is in line with the approach that we have undertaken in Ref. [3].

For a single layer graphene sheet of large size, Klemens found that graphene has "thermal conductivity larger than graphite by a factor of 2.3; its thermal conductivity is about $4.4 \times 10^3$ W/mK" [6]. For smaller graphene sheets, he determined that the thermal conductivity will be only slightly larger than that of graphite. At the same time, he performed his calculations with an average value of Gruneisen parameter $\gamma=2$. As we now know from the *ab initio* calculations, the Gruneisen parameter is substantially smaller for the relevant phonon branches over a wide range of phonon wave vectors $q$ [7]. If one uses, Klemens' formula (23) from Ref. [6] and a conservative average estimate of $\gamma=1.5$ (which is still higher than *ab initio* values for TA and comparable to values for LA branches), one gets for the RT thermal conductivity $K=$ 4114, 4596 and 5078 W/mK for the graphene flake dimensions of 5, 10, and 20 μm, respectively. This is in good agreement with both our calculations [3] and measurements [4-5] for the single layer graphene. If one takes the Gruneisen parameter from Kong *et al* [1] and Klemens' formula from Ref. [6] one obtains even higher values of graphene's thermal conductivity exceeding those of the experiments [4-5].

To sum up, Kong *et al* [1] used an unspecified truncation procedure for calculation of the Umklapp-limited thermal conductivity of the single layer graphene to obtain a result, which has no meaning. If no truncation were used ($\omega_C=0$), the Umklapp-limited thermal conductivity would have been infinite. The cut-off frequency for the single layer graphene cannot be introduced, even approximately, by analogy with the bulk graphite since the ZO' mode is absent and no strong coupling or heat transfer in cross-plane direction is possible. By using the "bulk" cut-off frequency, Kong *et al* [1] could not have obtained anything else but the result close to the bulk graphite value. The conclusion about the dependence of the thermal conductivity of graphene on the number of atomic layers is based entirely on the selection of $\omega_C$ and thus also unwarranted. The finite thermal conductivity without the introduction of the unphysical $\omega_C$ can be obtained for the single layer graphene only if one considers other scattering mechanisms as it was done by Klemens [7] or Nika *et al* [3].

## References

[1] B. D. Kong, S. Paul, M. Buongiorno Nardelli and K. W. Kim, "First principles calculations of lattice thermal conductivity in mono- and bi-layer graphene," arXiv: 0902.0642v1 (February 3, 2009).

[2] P.G. Klemens and D.F. Pedraza, *Carbon*, **32**, 735 (1994).

[3] D.L. Nika, E.P. Pokatilov, A.S. Askerov and A.A. Balandin, "Phonon thermal conduction in graphene: The role of Umklapp and edge roughness scattering," arXiv: 0812.0518v2; to appear in *Phys. Rev. B* (submitted to PRB in December 2008).

[4] A. A. Balandin, S. Ghosh, W. Bao, I. Calizo, D. Teweldebrhan, F. Miao, and C. N. Lau, *Nano Lett*., **8**, 902 (2008).

[5] S. Ghosh, I. Calizo, D. Teweldebrhan, E. P. Pokatilov, D. L. Nika, A. A. Balandin, W. Bao, F. Miao, and C. N. Lau, *Appl. Phys. Lett*., **92**, 151911 (2008).

[6] P.G. Klemens, *J. Wide Bandgap Materials*, **7**, 332 (2000).

[7] N. Mounet and N. Marzari, *Phys. Rev. B*, **71**, 205214 (2005).